\documentclass[conference]{IEEEtran}
\IEEEoverridecommandlockouts

\usepackage{cite}
\usepackage{amsmath,amssymb,amsfonts}
\usepackage[linesnumbered,ruled,vlined]{algorithm2e}
\usepackage{algorithmicx}
\usepackage{subcaption}  
\usepackage{booktabs}
\usepackage{graphicx}
\usepackage{tabularx}
\usepackage[noend]{algpseudocode}
\usepackage{textcomp}
\usepackage{xcolor}
\usepackage{xspace}
\usepackage{tcolorbox}
\usepackage{enumitem}
\usepackage[pdfborder={0 0 2},colorlinks=false,linkbordercolor=red,citebordercolor=green,urlbordercolor=cyan]{hyperref}
\def\BibTeX{{\rm B\kern-.05em{\sc i\kern-.025em b}\kern-.08em
    T\kern-.1667em\lower.7ex\hbox{E}\kern-.125emX}}

\newcommand*{\bench}{{\text{CrossPL}}\xspace}
    
\begin{document}

\title{CrossPL: Evaluating Large Language Models on Cross Programming Language Code Generation}

\author
{
\IEEEauthorblockN{1\textsuperscript{st} Zhanhang Xiong}
\IEEEauthorblockA{\textit{Zhejiang University} \\
zhanhangxiong@zju.edu.cn\\
Hangzhou, China}
\and
\IEEEauthorblockN{2\textsuperscript{nd} Dongxia Wang}
\IEEEauthorblockA{\textit{Zhejiang University} \\
dxwang@zju.edu.cn\\
Hangzhou, China}
\and
\IEEEauthorblockN{3\textsuperscript{rd} Yuekang Li}
\IEEEauthorblockA{\textit{The University of New South Wales} \\
yuekang.li@unsw.edu.au\\
Sydney, Australia}
\and
\IEEEauthorblockN{4\textsuperscript{st} Xinyuan An}
\IEEEauthorblockA{\textit{Zhejiang University} \\
anxy@zju.edu.cn\\
Hangzhou, China}
\and
\IEEEauthorblockN{5\textsuperscript{st} Wenhai Wang}
\IEEEauthorblockA{\textit{Zhejiang University} \\
zdzzlab@zju.edu.cn\\
Hangzhou, China}
}

\maketitle
\begin{abstract}
As large language models (LLMs) become increasingly embedded in software engineering workflows, a critical capability remains underexplored: generating correct code that enables cross-programming-language (CPL) interoperability. This skill is essential for building complex systems that integrate components written in multiple languages via mechanisms like inter-process communication (IPC). To bridge this gap, we present CrossPL, the first benchmark designed to systematically evaluate LLMs’ ability to generate CPL-interoperating code. CrossPL comprises 1,982 tasks centered around IPC, covering six widely-used programming languages and seven representative CPL techniques. We construct this benchmark by (i) analyzing 19,169 multi-language GitHub repositories using 156 hand-crafted finite state machines (FSMs), and (ii) developing an LLM-based pipeline that automatically extracts CPL code snippets, generates task instructions, and validates functional correctness. We evaluate 14 state-of-the-art general-purpose LLMs and 6 code-oriented LLMs released in the past three years on CrossPL via FSM-based validation. Results reveal that even the best-performing models struggle with CPL scenarios, underscoring the need for more targeted research in this space. Our benchmark and code are available at: \url{https://anonymous.4open.science/r/crosspl-2814/}.
\end{abstract}

\section{Introduction}

In software system development, the use of multiple programming languages (MPL) has become increasingly common, as it allows developers to exploit the unique strengths of different languages to improve performance, modularity, and scalability \cite{muti1, muti2, muti3, muti4, muti5}. However, MPL software systems also introduce significant complexity, particularly in ensuring correct and efficient interoperating across heterogeneous languages \cite{mutierror1, mutierror2, mutierror3, interface, interface2, interface3, interface4}.

Large language models (LLMs) have shown impressive capabilities in code generation and reasoning \cite{reason, starcoder2}, leading to widespread adoption and productivity gains in software development workflows \cite{llmsurvey}, offering new promise for the development of MPL software systems as well \cite{mutichat}. Nevertheless, a critical question remains unresolved: \textbf{Can LLMs accurately generate cross-programming language (CPL) interoperating code?}
\begin{figure}[t]
  \centering
  \includegraphics[width=0.5\textwidth]{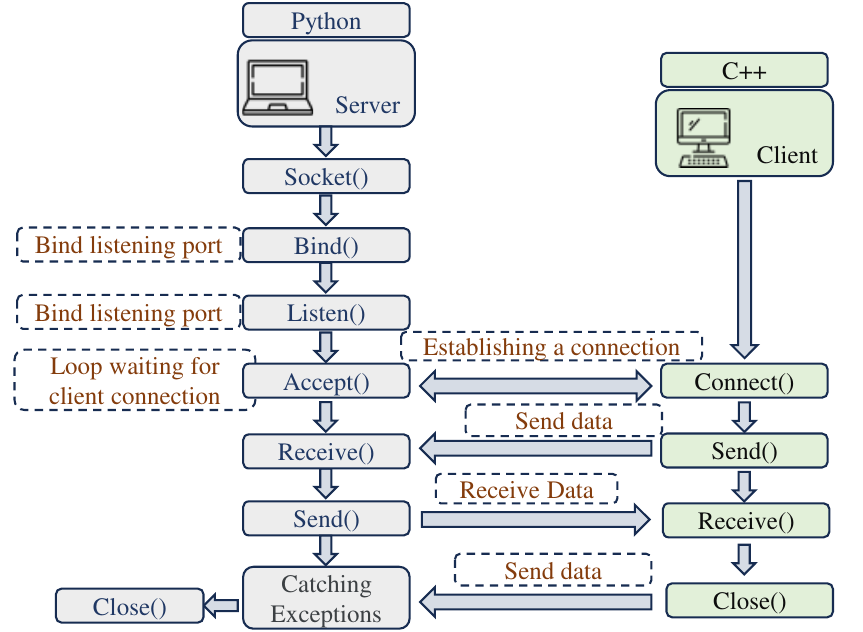}
  \caption{An example of IPC (Socket) between Python and C++.}
  \label{fig:demo_ipc}
\end{figure}

Existing benchmarks for evaluating the code generation capabilities of LLMs primarily focus on general-purpose code generation\cite{humaneval, mbpp, javabench, classeval, class2, repo1, repo2, repo3} and tasks in specific domains \cite{web1, web2, web3, dacode, dsbench, ds3, embed1, robot, domaineval}. Although some studies \cite{ huamanevalx, mutipl-e, cruxeval-x} have assessed LLMs' ability to generate MPL code, these studies typically involve translating tasks from one language to another, which does not evaluate LLMs' ability to generate code involving CPL interactions directly.
To the best of the authors’ knowledge, the current software engineering community lacks a dedicated benchmark for evaluating LLM’s ability to generate code that enables CPL interoperating. Addressing this gap is both timely and essential. A specialized benchmark would not only establish a foundation for systematic evaluation, but also catalyze future improvements in LLMs’ ability to support CPL and cross-platform software development.

MPL software systems typically rely on two primary mechanisms to enable interaction between components written in different programming languages: Inter-process Communication (IPC) and Foreign Function Interface (FFI) \cite{interface,poly1,polyfax}. IPC facilitates communication between independent processes, often using \textit{Sockets}, \textit{gRPC}, \textit{message queues}, or other cross-platform protocols. 
Although both FFI and IPC provide essential support for CPL interoperating, \bench focuses primarily on IPC-based interactions. This decision is driven by the fact that IPC is more widely used in real-world MPL systems, owing to its language-agnostic nature, decoupled architecture, and better support for distributed deployment. Fig. ~\ref{fig:demo_ipc} illustrates an example of cross-language interaction between Python and C++ by an IPC protocol (\textit{Socket}). Such examples are widely found in MPL projects involving Python and C++ for data science, robotics, and embedded systems

However, constructing a CPL interoperating benchmark centered on IPC also presents several challenges:
1) In real-world MPL software systems, CPL interaction code snippets are typically sparse and implicitly embedded within large codebases, making it difficult to accurately detect the boundaries of CPL communication. Manual collection is time-consuming and incurs substantial overhead.
2) IPC encompasses a wide range of mechanisms, including but not limited to \textit{pipe}, \textit{message queue}, \textit{gRPC}, \textit{TCP}, \textit{UDP}, \textit{HTTP} and \textit{Websocket}, each exhibiting distinct characteristics, usage patterns, and programming paradigms. This diversity adds complexity to interface categorization and identification.
3) Given the involvement of MPL in CPL interoperating, establishing standardized and comprehensive evaluation metrics that fairly assess CPL interaction remains a significant challenge.

To address the practical challenges in benchmark construction, we propose three core strategies: 1) We conduct comprehensive interface analysis by reviewing official documentation of mainstream IPC technologies and designing 156 scenario-specific finite state machines (FSMs) for real-world IPC pattern detection and evaluation; 2) To reduce manual workload and enhance scalability, we implement an automated LLM-based pipeline that extracts CPL code snippets and generates corresponding instructions; 3) During evaluation, we apply FSM-based validation tailored to each IPC scenario to validate the functional correctness of LLM-generated code.

Following these strategies, we curated 1982 high-quality CPL tasks from 19169 MPL projects in Github. 
The benchmark contains 615 \textit{Java}, 502 \textit{Python}, 271 \textit{JavaScript}, 151 \textit{PHP}, 392 \textit{Go}, and 51 \textit{C++} tasks. Regarding IPC technologies, the dataset includes 779 \textit{HTTP}, 433 \textit{TCP}, 92 \textit{UDP}, 198 \textit{WebSocket}, 153 \textit{Pipe}, 129 \textit{gRPC}, and 198 \textit{message queue} tasks.
Finally, to answer the key question: \textbf{Can LLMs accurately generate CPL interoperating code?} We raise three key research questions (RQs) as follows:
\begin{itemize}[leftmargin=*]
    \item \textbf{RQ1}: How effectively can LLMs generate IPC code across different programming languages in CPL projects?
    \item \textbf{RQ2}: How does the performance of LLMs vary when generating code for different IPC techniques?
    \item \textbf{RQ3}: Impact of Model Characteristics on the CPL interoperating code generation Performance of LLMs.
\end{itemize}

We systematically investigate these RQs and have the following key findings: 1) LLMs tend to underperform on most tasks involving different programming languages, and their capabilities vary significantly across different subsets of programming language; 2) Similarly, LLMs show limited effectiveness in generating code for various IPC techniques, with substantial differences in performance depending on the specific technique; 3) Recent advances in model thinking techniques do not significantly improve the accuracy of IPC code generation by LLMs and, in some cases, even under-perform compared to base models. These findings reveal the inadequacy of existing LLMs in generating CPL code, even though these models can perform well on single language coding tasks\cite{humaneval,mbpp}, highlighting the need for dedicated enhancement in this important code generation scenario.

We summarize our main contributions as follows:
\begin{itemize}[leftmargin=*]
\item \textbf{\bench benchmark:} We propose \bench{}, to our knowledge the first benchmark aimed at evaluating the ability of LLMs to generate CPL interoperating code involving IPC. It comprises 1982 instances, encompassing six programming languages and seven major IPC technologies.

\item \textbf{Comprehensive FSM-based interface characterization:} We carefully constructed 156 FSMs based on the official CPL interface specifications to formally characterize the IPC-based interaction interfaces. Such FSM-based characterization can not only facilitate us to detect IPC code snippets in real-world GitHub repositories, but also used to evaluate the capability of LLMs to generate CPL code under specific IPC scenarios. 

\item \textbf{LLM-based automatic analysis workflow:} Based on the FSM-based interaction characterization, we further develop a LLM-based workflow that automatically extract relevant CPL code snippets, generate natural-language prompts and construct evaluation tasks for constructing the benchmark.

\item \textbf{Large-scale empirical study:} We evaluate 20 representative LLMs to \textbf{answer the key question: whether LLMs can accurately generate cross-language interoperating code.} The findings highlight the need for more dedicated effort in this critical yet underexplored area.
\end{itemize}

The code and benchmark dataset for this project are publicly available at \url{https://anonymous.4open.science/r/crosspl-2814/}.
\begin{figure*}[htbp]
  \centering
  \includegraphics[width=0.92\textwidth]{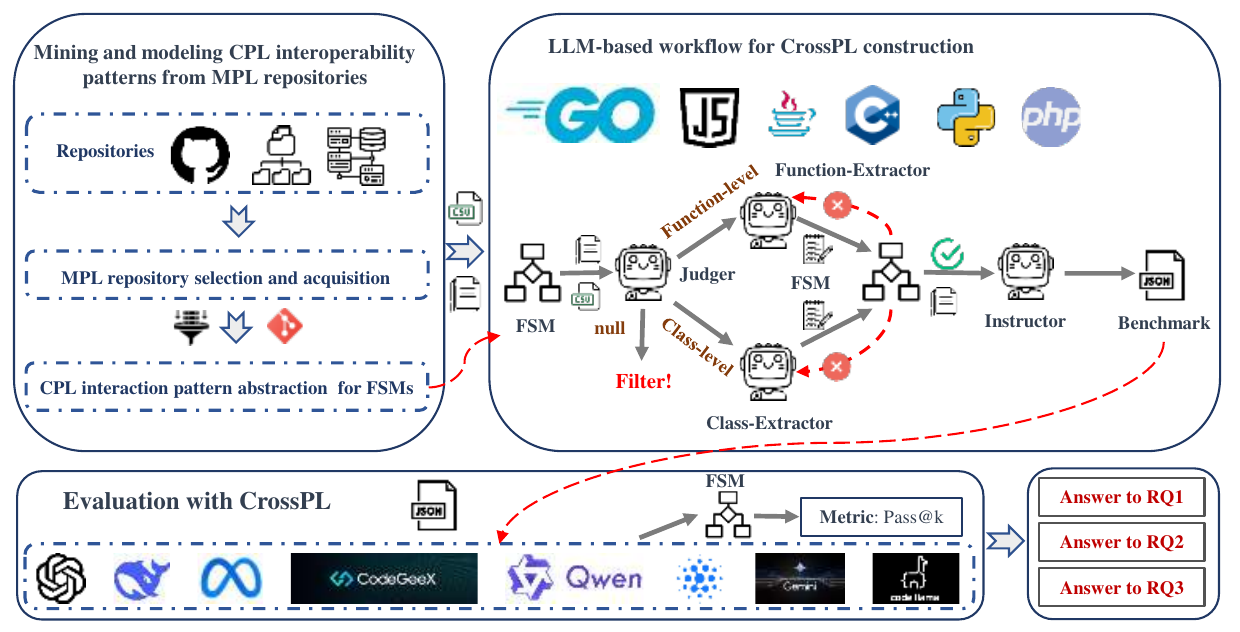}
  \caption{Framework for CPL Interoperating Code Analysis, Extraction, Generation and Evaluation.}
  \label{fig:framework}
\end{figure*}
\section{Related work}

\textbf{Single Language Code Generation Benchmarks.} Early benchmarks like BLEU \cite{bleu} and CodeBLEU \cite{codebleu} focus on surface or structural similarity but fail to capture functional correctness. HumanEval \cite{humaneval} introduces pass@k for functional evaluation and becomes standard. However, these benchmarks primarily target single-language, function-level tasks, mostly in Python. As evaluation expands, ClassEval \cite{classeval} and CodeAgentBench \cite{repo1} assess class-level and repository-level performance, still within Python. Domain-specific benchmarks, data science \cite{dacode, dsbench,ds3}, embedded systems \cite{embed1}, robotics \cite{robot}, and other domains \cite{domaineval}, also remain confined to a single language in their respective fields.

\textbf{Multilingual Code Generation Benchmarks.} While early efforts focus on single-language settings, recent work shifts toward multilingual code generation. Xu et al. \cite{huamanevalx} introduce HumanEval-X, extending evaluation to five programming languages with 820 test samples. Cassano et al. \cite{mutipl-e} propose MultiPL-E, translating HumanEval and MBPP tasks into 18 languages to analyze multilingual performance. CRUXEVAL-X \cite{cruxeval-x} further assesses LLMs’ reasoning and generation across languages. HumanEval-XL \cite{huamaneval-xl} expands this effort, linking 23 natural and 12 programming languages with 22,080 prompts to create a large-scale multilingual benchmark.

Although these benchmarks have advanced the evaluation of LLMs' code generation capabilities to some extent, a crucial open question remains on the path toward general intelligence for LLMs, namely \textbf{ whether LLMs can accurately and effectively generate CPL interoperating code}. To the best of the authors’ knowledge, there is currently no suitable benchmark in the LLM research community specifically designed to assess LLMs’ ability to generate CPL interaction code.
\section{Methodology}
This section outlines the methodology employed to construct the benchmark for evaluating CPL interoperating code generation. Fig.\ref{fig:framework} presents an overview of the core pipeline of this work, spanning from code analysis and collection to code generation and evaluation. The framework consists of three main components: (1) Mining and modeling CPL interoperating patterns from MPL repositories; (2) CPL interoperating code snippets extraction and \bench benchmark construction; (3) Code generation and evaluation with \bench. Algorithm~\ref{alg:cpl} presents the full workflow involved in the analysis, extraction, generation, and evaluation of CPL code.

\begin{algorithm}[htbp]
\footnotesize
\caption{CPL code analysis, extraction, generation and evaluation}
\label{alg:cpl}
\SetKwInOut{Input}{Input}
\SetKwInOut{Output}{Output}

\Input{$\omega_{\text{min}}, \omega_{\text{max}}$: project star count bounds; 
$Ln_{\text{min}}, Ln_{\text{max}}$: number of languages per project bounds; 
$\mathcal{P}$: all public GitHub repositories within bounds.
$\mathcal{FSM}$: A list of FSMs. 
}
\BlankLine
\Output{
    $True\_List$: LLM's correct task information list; $False\_List$: LLM's wrong task information list 
}
\BlankLine
\textbf{Step 1: MPL Repositories Crawling} 

$\mathcal{R} \gets \text{Crawler}(\mathcal{P}, \omega_{\text{min}}, \omega_{\text{max}}, Ln_{\text{min}}, Ln_{\text{max}})$

\text{save} $\mathcal{R}$ to \text{CSV} \Comment{\textcolor{blue}{Record metadata of satisfied MPL projects}}

\BlankLine
\textbf{Step 2: Clone to Local} 

\ForEach{$r_i \in \mathcal{R}$}{
    $local\_dir \gets  r_i.\text{name} + \_ + r_i.\text{ID}$ 
    
    \text{Clone}($r_i.\text{CloneUrl}$, $local\_dir$) 
    
    \Comment{\textcolor{blue}{Clone to local with specific directory name}}
}
\BlankLine
\textbf{Step 3: FSM-based Interoperability Analysis} 

$\mathcal{F} \gets \emptyset$

\ForEach{$r_i \in \mathcal{R}$}{

    \ForEach{$\mathcal{FSM}_j \in \mathcal{FSM}$ }{
    \If{$\mathcal{FSM}_j(r_i)$}{ 
        $(p_i, \tau_i, \theta_i, L_i, \sigma_i, K_i) \gets \mathcal{FSM}_j(r_i)$ 

        \Comment\textcolor{blue}{{Algorithm of FSM can find in \cite{polyfax}}}

        $\mathcal{F}.append((p_i, \tau_i, \theta_i, L_i, \sigma_i, K_i))$
            
        \textbf{break}
    }
    \Else{
    
        \textbf{continue}
    }
    }
    
}
\BlankLine

\textbf{Step 4: Benchmark Construction (Algorithm \ref{alg:ipc})}
\BlankLine
\textbf{Step 5: Benchmark Evaluation} 

\ForEach{$\mathcal{B}_i \in \mathcal{B}$}{

    $code \gets LLM (I_i)$
    
    $\mathcal{V} \gets \mathcal{FSM}(\sigma_i)$
    
    \If{$\mathcal{V}(code)$}{
        $True\_List.append(\mathcal{B}_i)$
    }
    \Else{
        $False\_List.append(\mathcal{B}_i)$
    }
}
\end{algorithm}

\subsection{Step 1: Mining and modeling CPL interoperating patterns from MPL repositories}
\noindent\textbf{Repository Selection and Localization for CPL Analysis.} We begin by curating 19169 multi-language (MPL) projects from GitHub, each with 1000-30000 stars and employing 2–5 programming languages. For each project repository, we record the key metadata $\mathcal{R}$ (e.g., ID, name, star count, language set, description, timestamps) and store it in a structured CSV format. Using the provided clone URLs, we locally replicate all the repositories via Git to enable downstream analysis and extraction of cross-language interoperating code.

\noindent\textbf{Modeling CPL interoperating Patterns.} The main advantages of FSMs in clear structure and logic make them well-suited for managing complex processes with distinct stages and transitions. They thus are particularly suitable for analyzing IPC workflows, where interactions typically involve well-defined states (e.g., initialization, data transmission, termination) and deterministic transitions between them.

\begin{figure*}[t]
  \centering
  \includegraphics[width=0.7\textwidth]{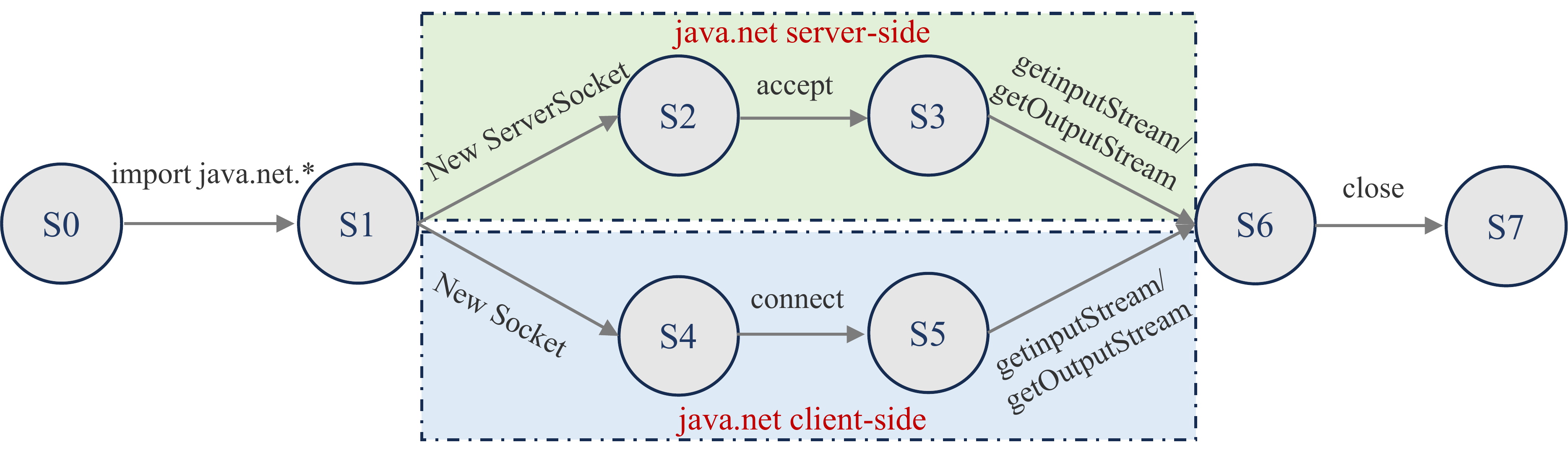}  
  \caption{An example of FSM-modeled CPL interoperating.}
  \label{fig:FSM}
\end{figure*}

Fig.~\ref{fig:FSM} illustrates an example of modeling CPL interoperating using FSMs, specifically modeling typical client-server interactions based on the \textit{java.net}. The upper part of the diagram models the server-side workflow, beginning with the import of networking libraries, followed by creating a \textit{ServerSocket}, accepting connections, establishing input or output streams, and eventually closing the \textit{socket}. The lower part represents the client-side process, including socket creation, connection initiation, stream handling, and termination. These workflows converge at the data exchange stage, representing the communication interface. This example demonstrates how FSMs can model API usage patterns in real-world code, enabling structured and interpretable modeling of communication logic.

PolyFax \cite{polyfax} first proposes leveraging FSMs to identify and classify interoperating patterns in MPL projects. However, it defines only 8 IPC-related FSMs, whose state patterns are overly coarse often with very few states and relying on fuzzy matching. This can lead to misclassification, especially since the FSMs do not account for code comments, making them prone to incorrectly matching commented-out code.

In this study, after analyzing extensive documentation and real-world MPL projects, we design a suite of 156 FSMs. Compared to that of PolyFax, our models offer far greater state and transition granularity, enabling accurate capture of each critical step in diverse CPL interoperating scenarios. Furthermore, our FSMs include three key enhancements:
\begin{enumerate}[leftmargin=*]
\item Adding semantic descriptions for each key step of CPL interoperating code, providing contextual semantics for the LLMs (like for instruction generation) in benchmark construction workflow.
\item Increasing the number of states to explicitly cover finer-grained IPC-specific steps, and designing clearer transition logic to reduce ambiguity in pattern detection.
\item Skipping code comments to avoid false positives from comment text matching critical step patterns.
\end{enumerate}

\subsection{Step 2: CPL interoperating code snippets extraction and benchmark construction}
Step 2 aims to identify CPL interoperating instances across the repositories using the 156 FSMs, generate their corresponding instructions and also to append the validation FSM of the corresponding task, for benchmark construction. For each matched instance, we record the file path $p$, interaction type $\tau$, interaction technique name $\theta$, programming language $L$, FSM identifier $\sigma$, and key procedural steps $K$. 
\begin{algorithm}[t]
\footnotesize
\caption{CPL Benchmark Construction}
\label{alg:ipc}

\SetKwInOut{Input}{Input}
\SetKwInOut{Output}{Output}
\Input{

$\mathcal{F} = \{(p_i, \tau_i, \theta_i, L_i, \sigma_i, K_i)\}_{i=1}^N$: metadata of CPL interoperating instances detected by FSMs; 

$N$: number of IPC instances identified by the FSMs;

$p_i$: file path of instance $i$; 

$\tau_i$: interaction type; 

$\theta_i$: technique used; 

$L_i$: programming language; 

$\sigma_i$: FSM ID; 

$K_i$: key steps of IPC.}
\BlankLine
\Output{IPC benchmark set $\mathcal{B}$ saved in JSON format.}
\BlankLine
$\mathrm{T_{id}} \gets 1$ \Comment{\textcolor{blue}{Initialize task identifier}}

\ForEach{$(p_i, \tau_i, \theta_i, L_i, \sigma_i, K_i) \in \mathcal{F}$}{
    $C_i \gets \text{read}(p_i)$\Comment{\textcolor{blue}{Read the code file of the instance}}
    
    $cl_i \gets \mathcal{A}_{1}(C_i, \tau_i, \theta_i, L_i, K_i)${}\Comment{\textcolor{blue}{Judger}}
    
    \If{$cl_i == \textit{null}$}{
        \textbf{continue}
    }
    \Else{
        
        $\mathcal{R}_i \gets \text{select\_reference}(C_i, cl_i)$

        \Comment{\textcolor{blue}{Select the required one-shot example}}

        \For{$k \gets 0$ \KwTo $5$}{
            
            $t \gets 0.1 \cdot k$
            \Comment{\textcolor{blue}{Set LLM temperature}}
            
            $c_k \gets \mathcal{A}_{2}(C_i, \tau_i, \theta_i, L_i, K_i, \mathcal{R}_i, cl_i, t)$

            \Comment{\textcolor{blue}{Code extracter}}

            \If{$\mathcal{A}_{3}(c_k, \sigma_i)$}{
                \Comment{\textcolor{blue}{FSM-based validation}}
                
                $I_k \gets \mathcal{A}_{4}(c_k, \tau_i, \theta_i, L_i, K_i)$

                \Comment{\textcolor{blue}{Instructor}}
                
                $\mathcal{B}_k \gets \{\mathrm{T_{id}}, p_i, \tau_i, \theta_i, L_i, \sigma_i, K_i, cl_i, I_k, c_k\}$

                Save$(\mathcal{B}_k, \text{save\_path}, \mathrm{T_{id}})$
                
                $\mathrm{T_{id}} \gets \mathrm{T_{id}} + 1$
        
                \textbf{break}
            }
        }
    }
}
\end{algorithm}

To reduce the manual load and the limitations of single LLMs in handling complex tasks with long contextual dependencies, we develop a LLM-based workflow powered by the DeepSeek-V3 \cite{deepseek} model\footnote{Selected for its state-of-the-art performance among open-source LLMs available prior to April 2025 and also its cost-effectiveness.} to construct \bench. Input code instances identified by FSMs as containing IPC techniques are first validated by the \textit{Judger} to confirm IPC implementation and classify them as function-level or class-level. Subsequently, instances are processed by the corresponding \textit{Function Extractor} or \textit{Class Extractor}, which extract minimal, logically complete IPC code snippets guided by technique-specific descriptions and one-shot examples. Extracted snippets undergo FSM-based validation, with up to five extraction attempts at increasing temperature if validation fails. Validated snippets are then passed to the LLM-based \textit{Instructor} to generate natural language task descriptions.  
Though currently a common practice\footnote{The feasibility of LLM-based code instruction generation has been shown in \cite{domaineval} and \cite{codebenchgen}.}, to ensure the effectiveness of such generated descriptions, we randomly sample and manually inspect a subset of them, confirming that they do clearly and accurately describe the intended tasks. Finally, the benchmark's task information, comprising metadata, code, and the generated instructions, are stored in structured JSON format. 

The entire workflow is formalized in Algorithm~\ref{alg:ipc}. 
Further details like the prompts used by each LLM are in~\href{https://anonymous.4open.science/r/crosspl-2814/}{here}.

\subsection{Step 3: Evaluation with \bench}
This final step involves generation and validation. We provide the instruction corresponding to each task in the benchmark to an LLM under evaluation, prompting it to generate the code for the respective functionality. Correctness is assessed by matching the generated code against the predefined FSMs to ensure protocol compliance. 
The use of FSMs is particularly advantageous here, as they not only offer a formal and modular way to represent the state transition logic of IPC technologies but also effectively capture subtle protocol violations or misuse patterns in the generated code.
\section{Statistics of \bench benchmark}
In constructing \bench benchmark, we conducted a thorough search and review of official documentation related to IPC technologies in CPL projects using keywords such as \textit{gRPC}, \textit{Pipe}, \textit{message queue}, \textit{TCP}, \textit{UDP}, \textit{WebSocket}, and \textit{HTTP} across different programming languages. 

\begin{figure}[t]
  \centering
  \begin{subfigure}[b]{0.5\textwidth}
    \includegraphics[width=\linewidth]{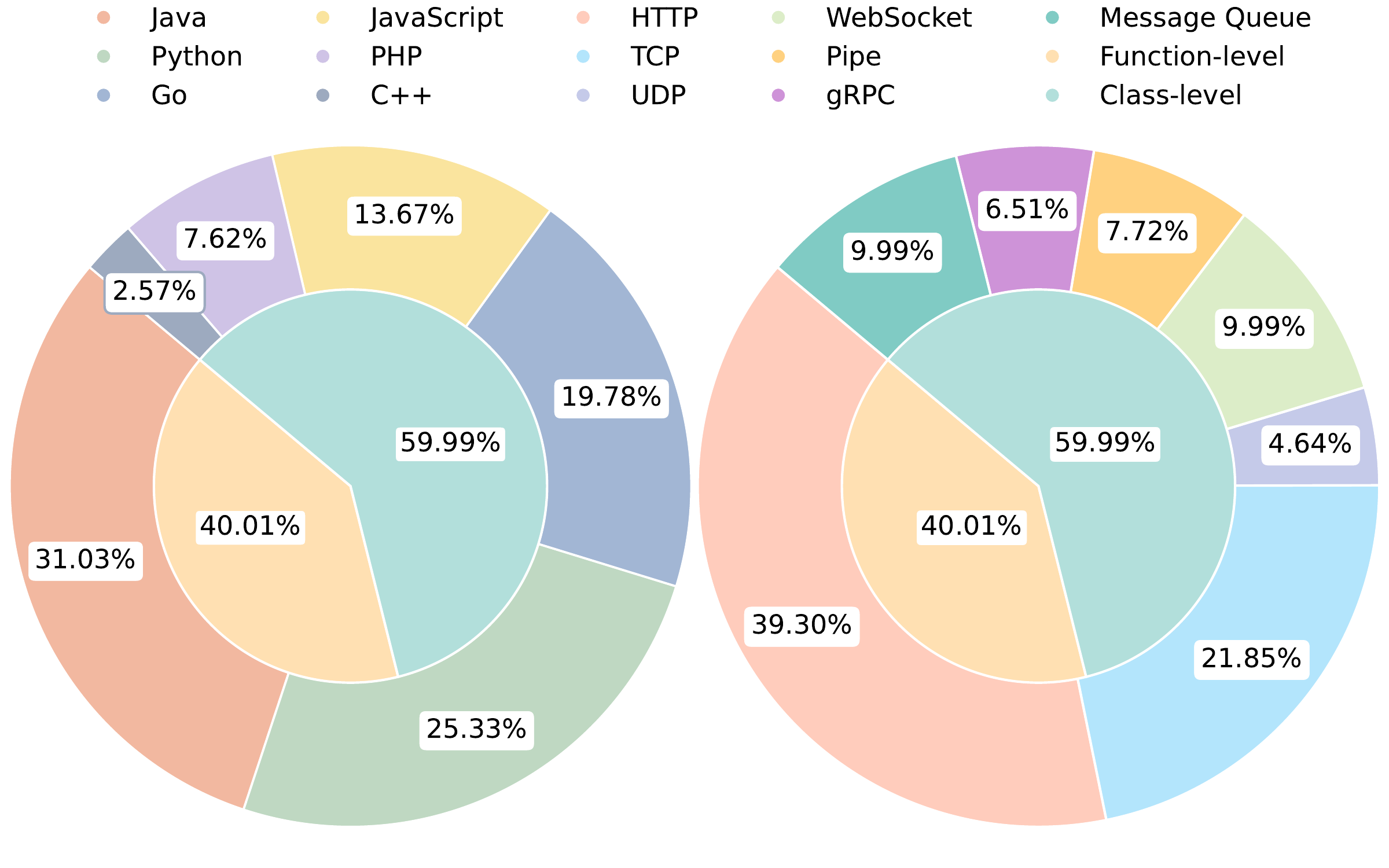}
    \caption{Distribution of \bench dataset from different view.}
    \label{fig3:a}
  \end{subfigure}
  \hfill
  \begin{subfigure}[b]{0.5\textwidth}
    \includegraphics[width=\linewidth]{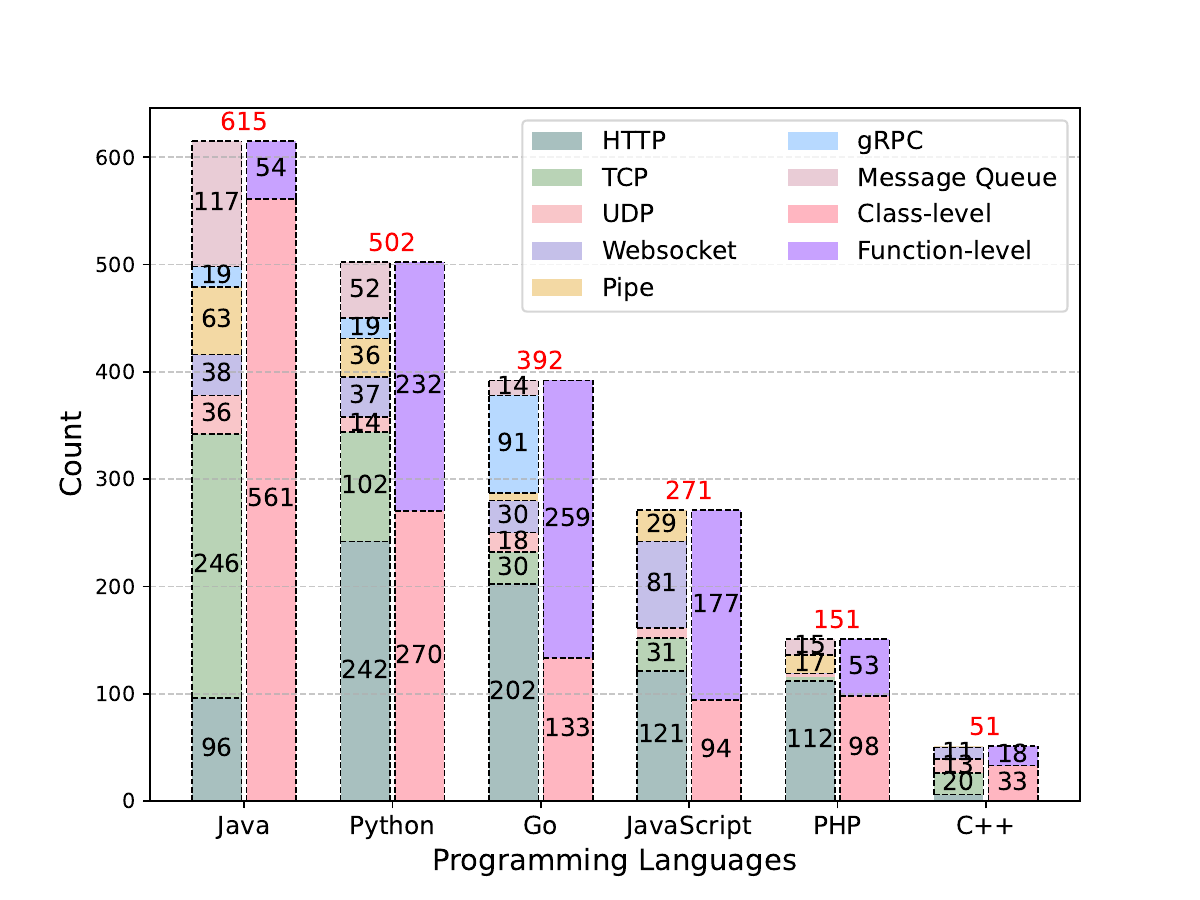}
    \caption{Distribution of different IPC technologies across different programming languages.}
    \label{fig3:b}
  \end{subfigure}
  \caption{Distribution of \bench benchmark.}
  \label{fig:doublewide}
\end{figure}

Fig.~\ref{fig3:a} summarizes the distribution in \bench from different perspectives. 
Overall it covers six programming languages and seven IPC technologies, comprising a total of 1982 tasks. 
Among the programming languages, Java accounts for the highest proportion of IPC-related tasks with 615 instances (31.03\%), whereas C++  the fewest, with 51 tasks (2.57\%). Among IPC technologies, \textit{HTTP} accounts for the highest proportion of tasks with 779 (39.30\%), while \textit{UDP} for the lowest with 92 tasks (4.64\%).

Fig.~\ref{fig3:b} further details the distribution of IPC technologies used by different programming languages within the \bench. Due to the distinct characteristics of each language, the proportional use of IPC technologies varies considerably; for example, Java tasks are predominantly associated with \textit{TCP}, while JavaScript tasks are mostly related to \textit{HTTP}. Notably, some IPC techniques cannot be fully implemented within a single function and require implementation at the class level. Consequently, during IPC-related code extraction, we distinguished between class-level and function-level code. Overall, class-level code accounts for 59.99\% of the \bench, while function-level code constitutes 40.01\%. This distribution reflects language-specific design patterns: for example, class-level implementations dominate in Java and PHP due to their object-oriented paradigms, while languages like Go and JavaScript tend to favor function-level or lightweight constructs. Such variations emphasize the necessity of handling both granularities in \bench to faithfully capture real-world IPC usage across different programming ecosystems.

\section{Experiment}

\begin{table*}[!t]
\renewcommand{\arraystretch}{1}
    \centering
    \caption{Performance of various models across programming languages (pass@1, greedy decode)}
    \resizebox{\textwidth}{!}{
    \begin{tabular}{clcccccccccc}
        \toprule
        Model Type & Model & Published Time & Size & Java ↑ & Python ↑ & Go ↑ & JavaScript ↑ & PHP ↑ & C++ ↑ & Mean ↑ \\ 
        \midrule
         & GPT-4o \cite{gpt-4o} & 2024 & $\backslash$ & 75.12\% & 67.13\% & 71.94\% & \textcolor{blue}{\underline{77.86\%}} & \textcolor{blue}{\underline{82.12\%}} & \textcolor{blue}{\underline{92.16\%}} & \textcolor{blue}{\underline{77.72\%}} \\ 
        ~ & GPT-4o-mini \cite{gpt-4o} & 2024 & $\backslash$ & 69.43\% & 62.15\% & 49.23\% & 73.80\% & 71.52\% & 80.39\% & 67.75\% \\ 
        ~ & Gemini-1.5-pro & 2024 & $\backslash$ & 72.52\% & 63.35\% & 66.33\% & 73.43\% & 78.15\% & 90.20\% & 74.00\% \\
        ~ & GLM4-plus \cite{glm4-plus} & 2024 & $\backslash$ & \textcolor{blue}{\underline{77.72\%}} & \textcolor{red}{\textbf{69.72\%}} & \textcolor{blue}{\underline{73.47\%}} & 79.34\% & \textcolor{red}{\textbf{82.12\%}} & \textcolor{red}{\textbf{96.08\%}} & \textcolor{red}{\textbf{79.74\%}} \\
        ~ & DeepSeek-V3 \cite{deepseek} & 2024 & 671B & \textcolor{red}{\textbf{78.21\%}} & 64.74\% & 66.84\% & \textcolor{red}{\textbf{82.66\% }}& 79.47\% & 88.24\% & 76.69\% \\
        ~ & Qwen3-235b-a22b \cite{qwen3} & 2025 & 235B & 77.40\% & \textcolor{blue}{\underline{67.93\%}} & 65.31\% & 76.38\% & 74.83\% & 88.24\% & 75.02\% \\ 
        ~ & Qwen3-32b \cite{qwen3} & 2025 & 32B & 70.89\% & 62.75\% & 49.74\% & 74.17\% & 75.50\% & 80.39\% & 68.91\% \\ 
        General Model & Qwen3-30b-a3b \cite{qwen3} & 2025 & 30B & 57.24\% & 51.99\% & 49.75\% & 70.48\% & 71.52\% & 66.67\% & 61.28\% \\ 
        ~ & Qwen3-14b \cite{qwen3} & 2025 & 14B & 67.64\% & 57.57\% & 59.44\% & 77.12\% & 76.82\% & 70.59\% & 68.20\% \\ 
        ~ & Qwen3-8b \cite{qwen3} & 2025 & 8B & 10.73\% & 55.78\% & 48.47\% & 68.63\% & 68.21\% & 64.71\% & 52.76\% \\ 
        ~ & Qwen3-4b \cite{qwen3} & 2025 & 4B & 54.96\% & 57.97\% & 52.81\% & 66.79\% & 57.62\% & 64.71\% & 59.14\% \\ 
        ~ & GLM4-9b \cite{glm} & 2024 & 9B & 65.04\% & 54.58\% & 56.12\% & 66.42\% & 70.86\% & 58.82\% & 61.97\% \\ 
        ~ & Lamma3-8b-instruct \cite{llama} & 2024 & 8B & 35.56\% & 45.42\% & 29.59\% & 53.87\% & 45.70\% & 31.37\% & 40.25\% \\ 
        ~ & Gemma-7b \cite{gemma} & 2024 & 7B & 52.52\% & 48.01\% & 39.80\% & 58.67\% & 64.90\% & 43.14\% & 51.17\% \\ 
        \midrule
         & CodeGeeX4 \cite{huamanevalx} & 2024 & $\backslash$ & 69.27\% & 62.55\% & 55.61\% & 69.00\% & 70.20\% & 80.39\% & 67.84\% \\ 
        ~ & Qwen2.5-coder-32b-instruct \cite{qwen2.5} & 2024 & 32B & 72.52\% & 65.74\% & \textcolor{red}{\textbf{73.98\%}} & 76.75\% & 77.48\% & 90.20\% & 76.11\% \\ 
        ~ & Qwen2.5-coder-14b-instruct \cite{qwen2.5} & 2024 & 14B & 70.41\% & 66.73\% & 69.90\% & 74.17\% & 78.81\% & 90.20\% & 75.04\% \\ 
        Code Model & Qwen2.5-coder-7b-instruct \cite{qwen2.5} & 2024 & 7B & 68.78\% & 63.55\% & 66.07\% & 75.65\% & 77.48\% & 84.31\% & 72.64\% \\ 
        ~ & CodeLlama-7b-instruct \cite{codellama} & 2023 & 7B & 53.33\% & 44.41\% & 49.74\% & 49.82\% & 54.30\% & 41.18\% & 48.80\% \\ 
        ~ & CodeGemma-7b \cite{codegemma} & 2024 & 7B & 52.20\% & 53.78\% & 52.30\% & 65.31\% & 62.91\% & 58.92\% & 57.57\% \\ 
        \bottomrule
    \end{tabular}}
    \label{tab:pass1}
\end{table*}

In order to answer the core question posed in this paper: \textbf{Can LLMs accurately generate code for CPL interactions?} We formulate three distinct research questions:
\begin{itemize}[leftmargin=*]
    \item \textbf{RQ1:} How effectively can LLMs generate IPC code across different programming languages in CPL projects?
    \item \textbf{RQ2:} How does the performance of LLMs vary when generating code for different IPC techniques in CPL projects?
    \item \textbf{RQ3:} How model characteristics influence CPL interoperating code generation performance of LLMs.
\end{itemize}

We also intended to compare \bench with similar benchmarks, but to the best of our knowledge, \bench is the first benchmark for CPL interaction, making such comparisons infeasible. Further details about \bench are available at \href{https://anonymous.4open.science/r/crosspl-2814/}{here}.

\subsection{Experimental settings}
In our evaluation, we adopt the unbiased \textit{Pass@k} metric \cite{humaneval} to precisely measure the functional correctness of code snippets generated by LLMs. Consistent with previous studies \cite{domaineval}, we report both \textit{Pass@1} and \textit{Pass@5} results under a zero-shot setting and use macro-averaging to calculate the overall scores. For \textit{Pass@1}, we employ greedy decoding by setting the temperature to 0. For \textit{Pass@5}, we select a minimum sample size of N = 5, with the temperature set to 0.2 and top-p sampling at 0.95.

For proprietary models and large-scale open-source models, we perform inference using their official APIs. For open-source models such as LLaMA3-8b-Instruct \cite{llama}, Gemma-7b \cite{gemma}, CodeLLaMA-7b-Instruct \cite{codellama}, and CodeGemma-7b \cite{codegemma}, we deploy them locally using Ollama on the NVIDIA RTX 3090 GPU.

\begin{table*}[!t]
\renewcommand{\arraystretch}{1}
    \centering
    \caption{Performance of various models across programming languages (pass@5, temperature=0.2, top-p=0.95)}
    \resizebox{\textwidth}{!}{
    \begin{tabular}{llcccccccccc}
        \toprule
        Model Type & Model & Published Time & Size & Java ↑ & Python ↑ & Go ↑ & JavaScript ↑ & PHP ↑ & C++ ↑ & Mean ↑ \\ 
        \midrule
         & GPT-4o & 2024 & / & 82.38\% & \textcolor{red}{\textbf{76.30\%}} & \textcolor{blue}{\underline{78.79\%}} & \textcolor{blue}{\underline{85.97\%}} & \textcolor{blue}{\underline{87.69\%}} & 89.76\% & \textcolor{red}{\textbf{83.48\%}} \\
        ~ & GPT-4o-mini  & 2024 & / & 74.16\% & 67.66\% & 54.01\% & 77.81\% & 73.45\% & 89.66\% & 72.79\% \\
        ~ & Gemini-1.5-pro  & 2024 & / & \textcolor{red}{\textbf{85.86\%}} & 68.32\% & 69.69\% & 77.39\% & 80.83\% & 87.75\% & 78.31\% \\
        ~ & GLM4-plus  & 2024 & / & 81.88\% & 72.74\% & 78.51\% & 82.27\% & 85.04\% & \textcolor{red}{\textbf{98.37\%}} & \textcolor{blue}{\underline{83.14\%}} \\
        ~ & DeepSeek-V3  & 2024 & 671B & 83.70\% & 72.82\% & 70.38\% & \textcolor{red}{\textbf{86.54\%}} & 83.13\% & 91.99\% & 81.43\% \\
        ~ & Qwen3-235b-a22b  & 2025 & 235B & \textcolor{blue}{\underline{84.69\%}} & \textcolor{blue}{\underline{76.16\%}} & 75.96\% & 82.67\% & 80.43\% & 93.64\% & 82.26\% \\
        ~ & Qwen3-32b  & 2025 & 32B & 79.73\% & 71.24\% & 77.77\% & 81.33\% & 80.13\% & 90.96\% & 80.19\% \\
        General Model & Qwen3-30b-a3b & 2025 & 30B & 63.89\% & 60.52\% & 55.66\% & 73.18\% & 76.21\% & 70.52\% & 66.66\% \\
        ~ & Qwen3-14b & 2025 & 14B & 74.22\% & 66.18\% & 64.44\% & 78.98\% & 80.04\% & 85.24\% & 74.85\% \\
        ~ & Qwen3-8b & 2025 & 8B & 13.56\% & 63.55\% & 54.88\% & 74.03\% & 77.07\% & 73.82\% & 59.49\% \\
        ~ & Qwen3-4b  & 2025 & 4B & 61.98\% & 63.09\% & 63.22\% & 72.19\% & 68.45\% & 68.15\% & 66.18\% \\
        ~ & GLM4-9b & 2024 & 9B & 77.06\% & 65.81\% & 67.30\% & 71.36\% & 75.97\% & 86.17\% & 73.95\% \\
        ~ & Lamma3-8b-instruct & 2024 & 70B & 63.54\% & 68.98\% & 54.81\% & 77.88\% & 67.54\% & 53.87\% & 64.44\% \\
        ~ & Gemma-7b  & 2024 & 7B & 71.23\% & 67.84\% & 60.98\% & 68.47\% & 77.07\% & 57.02\% & 67.10\% \\
        \midrule
         & CodeGeeX4  & 2024 & / & 76.93\% & 73.15\% & 66.95\% & 76.26\% & 79.27\% & 89.28\% & 74.51\% \\
        ~ & Qwen2.5-coder-32b-instruct & 2024 & 32B & 77.96\% & 71.95\% & 77.78\% & 81.56\% & 84.27\% & 94.07\% & 78.70\% \\
        ~ & Qwen2.5-coder-14b-instruct  & 2024 & 14B & 78.04\% & 74.40\% & 75.97\% & 81.41\% & 84.73\% & \textcolor{blue}{\underline{96.08\%}} & 78.91\% \\
        Code Model & Qwen2.5-coder-7b-instruct  & 2024 & 7B & 76.56\% & 74.06\% & 76.33\% & 80.34\% & \textcolor{red}{\textbf{87.98\%}} & 92.44\% & 79.05\% \\
        ~ & CodeLlama-7b-instruct  & 2023 & 7B & 80.88\% & 74.99\% & \textcolor{red}{\textbf{80.89\%}} & 74.02\% & 80.98\% & 86.56\% & 78.35\% \\
        ~ & CodeGemma-7b  & 2024 & 7B & 72.44\% & 70.86\% & 66.22\% & 73.50\% & 75.05\% & 74.31\% & 71.61\% \\
        \bottomrule
    \end{tabular}}
    \label{tab:pass5}
\end{table*}
\begin{figure}[htbp]
  \centering
    \includegraphics[width=\linewidth]{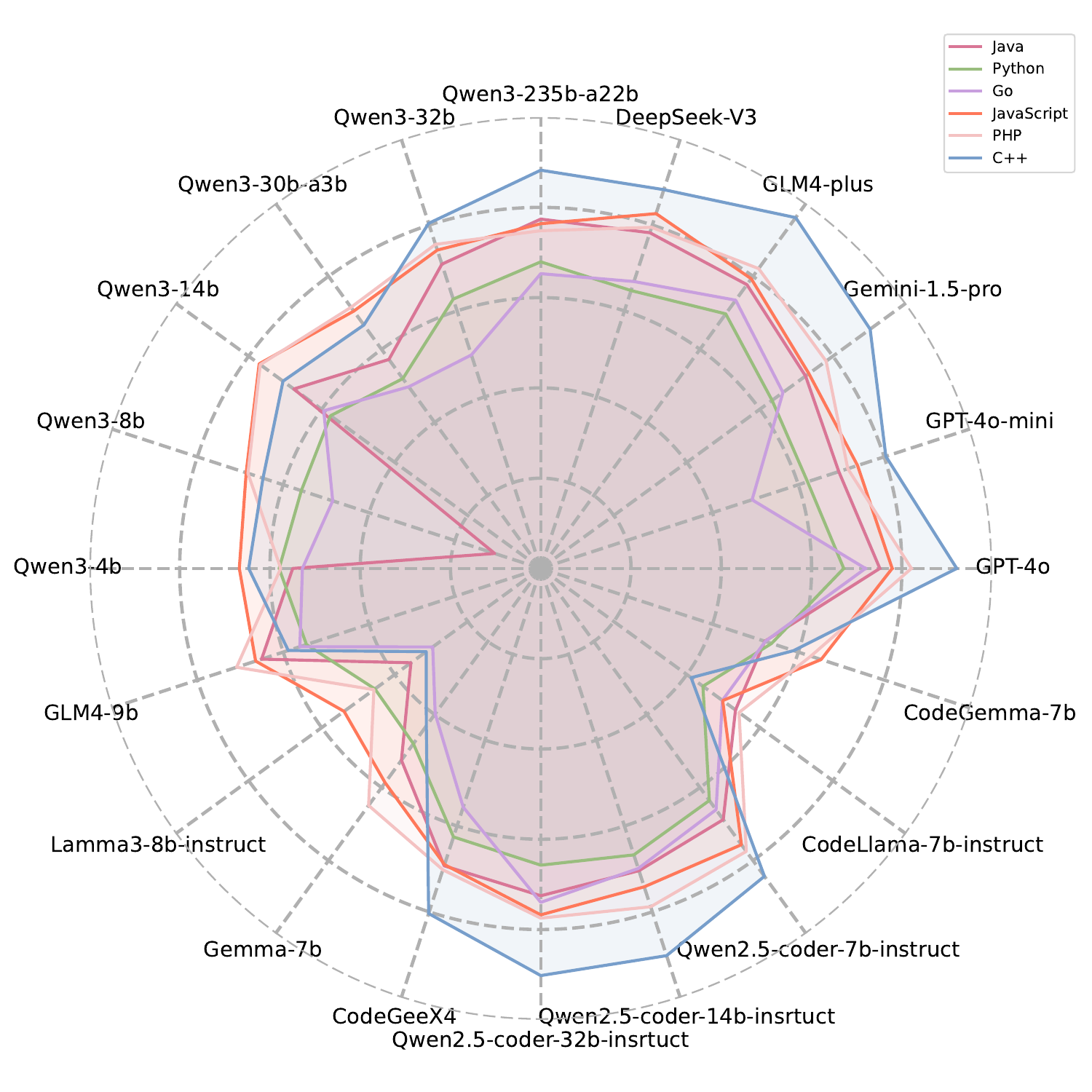}
    \caption{Radar Plot of Pass@1 across different programming languages}
    \label{fig:1a}
  \label{fig:ridar}
\end{figure}

\subsection{RQ1:How effectively can LLMs generate IPC code across
different programming languages in CPL projects?}

\textbf{Overall performance.} Table~\ref{tab:pass1} and Table~\ref{tab:pass5} present\footnote{Values highlighted in bold red indicate the best results, while those underlined in blue represent the second-best results. Similarly for the later tables.}  the \textit{pass@1} and \textit{pass@5} scores, respectively, for different LLMs across MPL on IPC tasks. Overall, \textit{pass@5} success rates consistently exceed their \textit{pass@1} counterparts across all the task subcategories, reflecting improved performance through multiple sampling attempts. Among the evaluated models, GLM4-plus achieves the highest average \textit{pass@1} score at 79.74\%, followed closely by GPT-4o at 77.72\%. In contrast, Qwen3-8b exhibits notably poor performance on Java IPC tasks, with a \textit{pass@1} of only 10.73\% and a \textit{pass@5} of 13.56\%.  

Further analysis of Table~\ref{tab:pass1} reveals that, among models with known parameter sizes, larger models generally exhibit better performance. However, this trend is not strictly linear. For instance, the 671b model (DeepSeek-V3) and the 235b model (Qwen3-235b-a22b) perform comparably, with an average difference of only 1.67\%. In contrast, both significantly outperform the 30b model (Qwen3-30b-a3b), with an average improvement of 14.66\%. Comparing Table~\ref{tab:pass1} with Table~\ref{tab:pass5}, pass@5 typically yields a substantial improvement over pass@1. However, some models with fewer than 10 billion parameters perform poorly, achieving pass@5 rates of only around 30\% or even lower.

\textbf{Effectiveness of code models.} 
Comparing general-purpose models with code-specific models, code models deliver comparable or even superior performance despite fewer parameters, emphasizing the effectiveness of LLMs' targeted domain adaptation. Analysis of Table~\ref{tab:pass1} indicates that, given comparable parameter sizes, code-specialized models tend to achieve higher performance. Taking the Go subset as an example, Qwen2.5-Coder-32B-Instruct outperforms Qwen3-32B by a substantial margin, with a 22.24\% higher pass@1 score.

\textbf{Impact of Code Structural Granularity on Model Performance.} An analysis of Tables~\ref{tab:pass1} and \ref{tab:pass5} reveals little correlation between model performance and code structural complexity. Although 91.22\% of candidate samples in the Java subset are at the class level (perceived more complex than the function level), the model does not exhibit the poorest performance on this subset. In contrast, the Go subset contains a higher proportion of function-level candidates (66.07\%), yet the model performs relatively poorly on it. 

\textbf{Language-specific performance disparities.} From the analysis, it is evident that the models perform best on C++ CPL tasks, while their performance on Go-related CPL tasks is relatively weaker. A potential explanation for this disparity lies in the structural characteristics of the Go language. Specifically, Go does not natively support the concept of classes, whereas much of the code knowledge learned by current LLMs is derived from class-based programming paradigms. In constructing class-level code samples for Go, we extracted code that emulates class-like behavior using structs. This structural mismatch may impose an additional cognitive burden on LLMs trained predominantly on class-based languages, thereby reducing their effectiveness when generating code in Go for such tasks. Conversely, the better performance on C++ tasks can be attributed to the CrossPL focus on widely used CPL techniques such as \textit{TCP}, \textit{UDP}, \textit{HTTP}, and \textit{WebSocket}. These protocols are well-documented and frequently appear in public code repositories, increasing the likelihood of LLM pretraining exposure. In addition, C++ is closely associated with low-level system programming, which aligns well with the structural characteristics of typical CPL implementations. Fig.~\ref{fig:ridar} illustrates the distribution of pass@1 scores across different programming languages. The figure clearly highlights performance discrepancies among tasks in various languages. Overall, the models achieve the best performance on C++, while their performance on Go is relatively weaker.

\begin{tcolorbox}[title= Answer to RQ1, boxrule=0.5pt, boxsep=0.5pt, left=1pt, right=1pt, top=0.5pt, bottom=0.5pt]
LLMs exhibit varying effectiveness in generating IPC code across programming languages. Performance is highest in C++, likely due to its alignment with well-documented IPC patterns and low-level system semantics. In contrast, performance declines in Go, potentially due to structural mismatches with class-oriented training data. Notably, in this experement, code structural granularity (class-level vs. function-level) alone does not appear to significantly influence model performance.
\end{tcolorbox}

\subsection{RQ2: How does the performance of LLMs vary when generating code for different IPC techniques?}

\begin{table*}[htbp]
\renewcommand{\arraystretch}{1}
    \centering
    \caption{Performance of various models across IPC techniques (pass@1, greedy decode)}
    \resizebox{\textwidth}{!}{
    \begin{tabular}{llcccccccccc}
        \toprule
        Model Type & Model & Published Time & Size & HTTP ↑ & TCP ↑ & UDP ↑ & Websocket ↑ & Pipe ↑ & gRPC ↑ & Message Queue ↑ & Mean ↑ \\ 
        \midrule
         & GPT-4o & 2024 & / & \textcolor{red}{\textbf{71.63\%}} & 71.13\% & 82.61\% & 82.32\% & 64.05\% & 86.82\% & 74.75\% & 76.19\% \\
        ~ & GPT-4o-mini & 2024 & / & 59.69\% & 68.13\% & 69.57\% & 72.73\% & 66.67\% & 59.69\% & 67.68\% & 66.31\% \\
        ~ & Gemini-1.5-pro & 2024 & / & 61.49\% & 74.60\% & 73.91\% & 79.29\% & \textcolor{red}{\textbf{68.63\%}} & 86.82\% & 72.22\% & 73.85\% \\
        ~ & GLM4-plus & 2024 & / & \textcolor{blue}{\underline{68.68\%}} & \textcolor{blue}{\underline{77.83\%}} & \textcolor{red}{\textbf{84.78\%}} & \textcolor{red}{\textbf{88.38\%}} & \textcolor{blue}{\underline{67.32\%}} & \textcolor{blue}{\underline{89.92\%}} & \textcolor{red}{\textbf{80.81\%}} & \textcolor{red}{\textbf{79.67\%}} \\
        ~ & DeepSeek-V3 & 2024 & 671B & 66.24\% & \textcolor{red}{\textbf{78.79\%}} & \textcolor{red}{\textbf{84.78\%}} & 82.83\% & 60.78\% & 87.60\% & \textcolor{blue}{\underline{76.26\% }}& \textcolor{blue}{\underline{76.75\%}} \\
        ~ & Qwen3-235b-a22b & 2025 & 235B & 67.01\% & 77.60\% & \textcolor{blue}{\underline{83.70\%}}& 74.75\% & 65.36\% & 86.82\% & 72.22\% & 75.35\% \\
        ~ & Qwen3-32b & 2025 & 32B & 66.20\% & 69.48\% & 77.03\% & 84.52\% & 54.11\% & 81.58\% & 73.91\% & 72.40\% \\
        General Model & Qwen3-30b-a3b & 2025 & 30B & 52.76\% & 64.43\% & 60.87\% & 62.63\% & 56.86\% & 56.59\% & 56.06\% & 58.60\% \\
        ~ & Qwen3-14b & 2025 & 14B & 59.44\% & 72.98\% & 70.65\% & 72.73\% & 51.63\% & 76.74\% & 67.17\% & 67.33\% \\
        ~ & Qwen3-8b & 2025 & 8B & 49.55\% & 29.79\% & 35.87\% & 52.53\% & 35.95\% & 62.79\% & 35.35\% & 43.12\% \\
        ~ & Qwen3-4b & 2025 & 4B & 54.43\% & 62.12\% & 63.04\% & 54.04\% & 48.37\% & 68.22\% & 59.09\% & 58.47\% \\
        ~ & GLM4-9b & 2024 & 9B & 56.74\% & 60.28\% & 60.87\% & 60.61\% & 60.13\% & 81.40\% & 68.18\% & 64.03\% \\
        ~ & Lamma3-8b-instruct & 2024 & 8B & 40.31\% & 44.80\% & 36.96\% & 44.44\% & 33.91\% & 38.76\% & 34.34\% & 39.07\% \\
        ~ & Gemma-7b & 2024 & 7B & 47.50\% & 56.12\% & 53.26\% & 49.49\% & 47.06\% & 52.71\% & 50.00\% & 50.88\% \\
        \midrule
         & CodeGeeX4 & 2024 & / & 60.08\% & 69.05\% & 66.30\% & 68.69\% & 62.75\% & 72.87\% & 69.70\% & 67.06\% \\
        ~ & Qwen2.5-coder-32b-instruct & 2024 & 32B & 67.27\% & 72.52\% & 78.26\% & \textcolor{blue}{\underline{84.34\%}} & 63.40\% & \textcolor{red}{\textbf{90.70\%}} & 73.74\% & 75.75\% \\
        ~ & Qwen2.5-coder-14b-instruct & 2024 & 14B & 65.47\% & 76.21\% & 67.39\% & 82.32\% & 62.09\% & 85.27\% & 69.70\% & 72.64\% \\
        Code Model & Qwen2.5-coder-7b-instruct & 2024 & 7B & 64.31\% & 71.36\% & 70.65\% & 75.25\% & 62.09\% & 81.40\% & 71.72\% & 70.97\% \\
        ~ & CodeLlama-7b-instruct & 2023 & 7B & 46.98\% & 57.74\% & 55.43\% & 44.95\% & 45.10\% & 60.47\% & 45.96\% & 50.95\% \\
        ~ & CodeGemma-7b & 2024 & 7B & 50.83\% & 63.05\% & 61.96\% & 62.63\% & 45.75\% & 54.26\% & 54.55\% & 56.15\% \\
        \bottomrule
    \end{tabular}}
    \label{tab:pass1_ipc}
\end{table*}

\begin{table*}[htbp]
\renewcommand{\arraystretch}{1}
    \centering
    \caption{Performance of various models across IPC techniques (pass@5, temperature=0.2, top-p=0.95)}
    \resizebox{\textwidth}{!}{
    \begin{tabular}{llcccccccccc}
        \toprule
        Model Type & Model & Published Time & Size & HTTP ↑ & TCP ↑ & UDP ↑ & Websocket ↑ & Pipe ↑ & gRPC ↑ & Message Queue ↑ & Mean ↑ \\ 
        \midrule
         & GPT-4o & 2024 & / & \textcolor{red}{\textbf{75.58\%}} & 81.66\% & 85.35\% & \textcolor{red}{\textbf{88.96\%}} & \textcolor{blue}{\underline{75.40\%}} & 91.75\% & \textcolor{blue}{\underline{82.74\%}} & \textcolor{red}{\textbf{83.06\%}} \\
         & GPT-4o-mini & 2024 & / & 62.95\% & 72.16\% & 75.03\% & 77.93\% & 69.92\% & 66.39\% & 72.53\% & 70.99\% \\
         & Gemini-1.5-pro & 2024 & / & 67.20\% & 83.42\% & 82.04\% & 81.77\% & 73.36\% & \textcolor{blue}{\underline{92.02\%}} & 77.89\% & 79.67\% \\
         & GLM4-plus & 2024 & / & \textcolor{blue}{\underline{71.47\%}} & 81.42\% & 86.84\% & 80.18\% & 71.20\% & \textcolor{red}{\textbf{93.17\%}} & \textcolor{red}{\textbf{84.25\%}}& 81.22\% \\
         & DeepSeek-V3 & 2024 & 671B & 70.06\% & \textcolor{blue}{\underline{84.92\%}} & \textcolor{blue}{\underline{87.99\%}} & 86.91\% & 68.58\% & 91.20\% & 80.84\% & 81.50\% \\
         & Qwen3-235b-a22b & 2025 & 235B & 73.80\% & \textcolor{red}{\textbf{85.55\%}} & \textcolor{red}{\textbf{88.62\%}} & 82.81\% & \textcolor{red}{\textbf{79.30\%}} & 89.49\% & 76.61\% & \textcolor{blue}{\underline{82.31\%}} \\
         & Qwen3-32b & 2025 & 32B & 72.12\% & 80.70\% & 81.68\% & 88.14\% & 65.45\% & 86.00\% & 77.88\% & 78.85\% \\
    General Model & Qwen3-30b-a3b & 2025 & 30B & 57.73\% & 70.75\% & 63.62\% & 67.08\% & 64.04\% & 63.84\% & 63.13\% & 64.31\% \\
         & Qwen3-14b & 2025 & 14B & 62.68\% & 78.91\% & 76.56\% & 79.18\% & 66.53\% & 81.29\% & 72.37\% & 73.93\% \\
         & Qwen3-8b & 2025 & 8B & 53.91\% & 36.66\% & 39.25\% & 58.91\% & 44.23\% & 68.85\% & 37.08\% & 48.41\% \\
         & Qwen3-4b & 2025 & 4B & 59.93\% & 67.66\% & 69.06\% & 61.24\% & 58.24\% & 80.28\% & 65.70\% & 66.02\% \\
         & GLM4-9b & 2024 & 9B & 63.63\% & 72.22\% & 67.93\% & 73.03\% & 72.46\% & 89.40\% & 76.52\% & 73.60\% \\
         & Lamma3-8b-instruct & 2024 & 8B & 59.75\% & 73.19\% & 62.68\% & 59.85\% & 62.28\% & 80.30\% & 62.14\% & 65.74\% \\
         & Gemma-7b & 2024 & 7B & 43.31\% & 57.26\% & 60.24\% & 51.23\% & 52.28\% & 65.82\% & 46.68\% & 53.83\% \\
        \midrule
         & CodeGeeX4 & 2024 & / & 68.49\% & 79.62\% & 76.58\% & 73.73\% & 73.73\% & 80.32\% & 76.04\% & 75.50\% \\
         & Qwen2.5-coder-32b-instruct & 2024 & 32B & 71.07\% & 79.52\% & 79.56\% & \textcolor{blue}{\underline{88.49\%}} & 73.13\% & 95.18\% & 77.72\% & 80.67\% \\
    Code Model & Qwen2.5-coder-14b-instruct & 2024 & 14B & 71.07\% & 82.77\% & 77.87\% & 87.11\% & 69.81\% & 91.35\% & 77.69\% & 79.67\% \\
         & Qwen2.5-coder-7b-instruct & 2024 & 7B & 72.75\% & 78.32\% & 79.16\% & 83.85\% & 67.12\% & 89.81\% & 81.73\% & 78.96\% \\
         & CodeLlama-7b-instruct & 2023 & 7B & 50.41\% & 66.99\% & 74.54\% & 63.97\% & 54.68\% & 78.20\% & 55.02\% & 63.40\% \\
         & CodeGemma-7b & 2024 & 7B & 64.38\% & 79.31\% & 75.47\% & 78.09\% & 65.98\% & 75.03\% & 74.63\% & 73.27\% \\
        \bottomrule
    \end{tabular}}
    \label{tab:pass5_ipc}
\end{table*}

\textbf{Overall performance.} Observe from Tables~\ref{tab:pass1_ipc} and~\ref{tab:pass5_ipc} that GLM4-plus achieves the best overall performance across the subsets of various IPC techniques. It attains the highest \textit{pass@1} scores in the \textit{UDP}, \textit{WebSocket}, and \textit{Message Queue} subsets, and ranks second-best in the remaining four subsets. In contrast, Llama3-8b-instruct exhibits the lowest average \textit{pass@1} score at only 39.07\%, while Qwen3-8b records the lowest average \textit{pass@5} score. 

\textbf{Notable outliers and analysis.} Interestingly, Gemma-7b exhibits lower \textit{pass@5} than \textit{pass@1} on the \textit{HTTP} and \textit{Message Queue} subsets, indicating that sampling introduces noise rather than diversity. Across other IPC subsets, limited \textit{pass@5} improvements further reveal Gemma-7b's difficulty in generating diverse yet correct alternatives, highlighting its reliance on deterministic decoding.

\textbf{IPC-specific performance disparities}: Tables~\ref{tab:pass1_ipc}, \ref{tab:pass5_ipc}, reveal significant variation in LLM performance across IPC subsets for CPL interoperating code. The \textit{gRPC} subset achieves the highest results, with 10 LLMs surpassing 80\% on pass@1 and 14 on pass@5; the top model reaches 90.70\% and 93.17\%, respectively. This success likely stems from \textit{gRPC}’s standardized syntax, schema-driven design, and extensive representation in training data. In contrast, the low-level, platform-dependent \textit{Pipe} subset underperforms, with 9 LLMs below 60\% on pass@1 and 10 below 70\% on pass@5; minimum scores are 33.91\% and 44.23\%. Similarly, the \textit{HTTP} subset shows weaker results—10 LLMs under 60\% pass@1 and 14 under 70\% pass@5, likely due to its variable implementations and lack of strict interfaces, hindering LLM generalization.

\begin{tcolorbox}[title= Answer to RQ2, boxrule=0.5pt, boxsep=0.5pt, left=1pt, right=1pt, top=1pt, bottom=1pt]
LLM performance varies across different IPC techniques. Higher-level protocols such as \textit{gRPC} tend to yield better results, likely because of their well-defined and structured semantics. In contrast, lower-level mechanisms like \textit{Pipe} present greater challenges, possibly due to their minimal abstraction and increased system-level complexity. These observations suggest that effective IPC code generation by LLMs may benefit from a deeper understanding of both protocol semantics and the underlying coordination logic.
\setlength{\tabcolsep}{1pt}
\end{tcolorbox}

\subsection{RQ3: Impact of Model Characteristics on the CPL interoperating code generation Performance of LLMs.}

\begin{table*}[!ht]
\renewcommand{\arraystretch}{1}
    \centering
    \caption{pass@1 results across Qwen3 series models on IPC-related tasks}
    \resizebox{\textwidth}{!}{
    \begin{tabular}{lccccccccccc}
        \toprule
        Model & Published Time & Size & Java ↑ & Python ↑ & Go ↑ & JavaScript ↑ & PHP ↑ & C++ ↑ & Mean ↑ \\ 
        \midrule
        Qwen3-235b-a22b-thinking & 2025 & 235B & \underline{\textcolor{blue}{75.12\%}} & \underline{\textcolor{blue}{67.73\%}} & \underline{\textcolor{blue}{61.48\%}} & 76.01\% & \underline{\textcolor{blue}{74.83\%}} & \underline{\textcolor{blue}{82.35\%}} & \underline{\textcolor{blue}{72.92\%}} \\
        Qwen3-32b-thinking       & 2025 & 32B  & 61.14\% & 60.36\% & 60.97\% & 72.32\% & 72.19\% & \underline{\textcolor{blue}{82.35\%}} & 68.22\% \\
        Qwen3-30b-a3b-thinking   & 2025 & 30B  & 50.33\% & 55.18\% & 52.04\% & 70.85\% & 73.51\% & 78.43\% & 63.39\% \\
        Qwen3-14b-thinking       & 2025 & 14B  & 57.07\% & 62.75\% & 58.67\% & 69.74\% & 68.87\% & 72.55\% & 64.94\% \\
        Qwen3-8b-thinking        & 2025 & 8B   & 24.72\% & 55.58\% & 46.94\% & 69.74\% & 70.20\% & 66.67\% & 55.64\% \\
        Qwen3-4b-thinking        & 2025 & 4B   & 33.82\% & 57.77\% & 50.51\% & 61.62\% & 52.32\% & 70.59\% & 54.44\% \\
        \midrule
        Qwen3-235b-a22b & 2025 & 235B & \textbf{\textcolor{red}{77.40\%}} & \textbf{\textcolor{red}{67.93\%}} & \textbf{\textcolor{red}{65.31\%}} & \textcolor{blue}{\underline{76.38\%}} & \underline{\textcolor{blue}{74.83\%}} & \textbf{\textcolor{red}{88.24\%}} & \textbf{\textcolor{red}{75.02\%}} \\
        Qwen3-32b & 2025 & 32B  & 70.89\% & 62.75\% & 49.74\% & 74.17\% & 75.50\% & 80.39\% & 68.91\% \\
        Qwen3-30b-a3b& 2025 & 30B  & 57.24\% & 51.99\% & 49.75\% & 70.48\% & 71.52\% & 66.67\% & 61.28\% \\
        Qwen3-14b & 2025 & 14B  & 67.64\% & 57.57\% & 59.44\% & \textbf{\textcolor{red}{77.12\%}} & \textbf{\textcolor{red}{76.82\%}} & 70.59\% & 68.20\% \\
        Qwen3-8b & 2025 & 8B   & 10.73\% & 55.78\% & 48.47\% & 68.63\% & 68.21\% & 64.71\% & 52.76\% \\
        Qwen3-4b & 2025 & 4B   & 54.96\% & 57.97\% & 52.81\% & 66.79\% & 57.62\% & 64.71\% & 59.14\% \\
        \bottomrule
    \end{tabular}
    }
    \label{tab:qwen3_performance}
\end{table*}

\begin{table*}[htbp]
\renewcommand{\arraystretch}{1}
    \centering
    \caption{Performance of Qwen3 series models across network protocols (pass@1)}
    \resizebox{\textwidth}{!}{
    \begin{tabular}{lcccccccccc}
        \toprule
         Model & Published Time & Size & HTTP ↑ & TCP ↑ & UDP ↑ & Websocket ↑ & Pipe ↑ & gRPC ↑ & Message Queue ↑ & Mean ↑ \\ 
        \midrule
          Qwen3-235b-a22b-think & 2025 & 235B & 64.96\% & \textcolor{blue}{\underline{76.67\%}} & \textcolor{blue}{\underline{79.35\%}} & \textcolor{blue}{\underline{73.74\%}} & \textcolor{blue}{\underline{64.05\%}} & \textcolor{blue}{\underline{82.17\%}} & \textcolor{blue}{\underline{72.22\%}} & \textcolor{blue}{\underline{73.31\%}} \\
          Qwen3-32b-think & 2025 & 32B & 58.28\% & 70.44\% & 70.65\% & 71.72\% & 45.10\% & 80.62\% & 63.64\% & 65.78\% \\
          Qwen3-30b-a3b-think & 2025 & 30B & 56.10\% & 60.28\% & 65.22\% & 64.14\% & 48.37\% & 66.67\% & 54.04\% & 59.26\% \\
          Qwen3-14b-think & 2025 & 14B & 55.58\% & 66.51\% & 58.70\% & 72.73\% & 54.25\% & 79.07\% & 61.62\% & 64.07\% \\
          Qwen3-8b-think & 2025 & 8B & 50.19\% & 41.80\% & 45.65\% & 59.60\% & 35.29\% & 54.26\% & 44.44\% & 47.32\% \\
          Qwen3-4b-think & 2025 & 4B & 51.86\% & 49.19\% & 43.48\% & 50.51\% & 32.68\% & 61.24\% & 46.46\% & 47.92\% \\
         \midrule
          Qwen3-235b-a22b & 2025 & 235B & \textcolor{red}{\textbf{67.01\%}} & \textcolor{red}{\textbf{77.60\%}} & \textcolor{red}{\textbf{83.70\%}} & 74.75\% & \textcolor{red}{\textbf{65.36\%}} & \textcolor{red}{\textbf{86.82\%}} & \textcolor{blue}{\underline{72.22\%}} & \textcolor{red}{\textbf{75.35\%}} \\
          Qwen3-32b & 2025 & 32B & \textcolor{blue}{\underline{66.20\%}}& 69.48\% & 77.03\% & \textcolor{red}{\textbf{84.52\%}}& 54.11\% & 81.58\% & \textcolor{red}{\textbf{73.91\%}} & 72.40\% \\
          Qwen3-30b-a3b & 2025 & 30B & 52.76\% & 64.43\% & 60.87\% & 62.63\% & 56.86\% & 56.59\% & 56.06\% & 58.60\% \\
          Qwen3-14b & 2025 & 14B & 59.44\% & 72.98\% & 70.65\% & 72.73\% & 51.63\% & 76.74\% & 67.17\% & 67.33\% \\
          Qwen3-8b & 2025 & 8B & 49.55\% & 29.79\% & 35.87\% & 52.53\% & 35.95\% & 62.79\% & 35.35\% & 43.12\% \\
          Qwen3-4b & 2025 & 4B & 54.43\% & 62.12\% & 63.04\% & 54.04\% & 48.37\% & 68.22\% & 59.09\% & 58.47\% \\
        \bottomrule
        \label{tab:qwen3_performance2}
    \end{tabular}}
\end{table*}
\begin{figure*}[t]
  \centering
  \begin{subfigure}[b]{0.49\textwidth}
    \includegraphics[width=\linewidth]{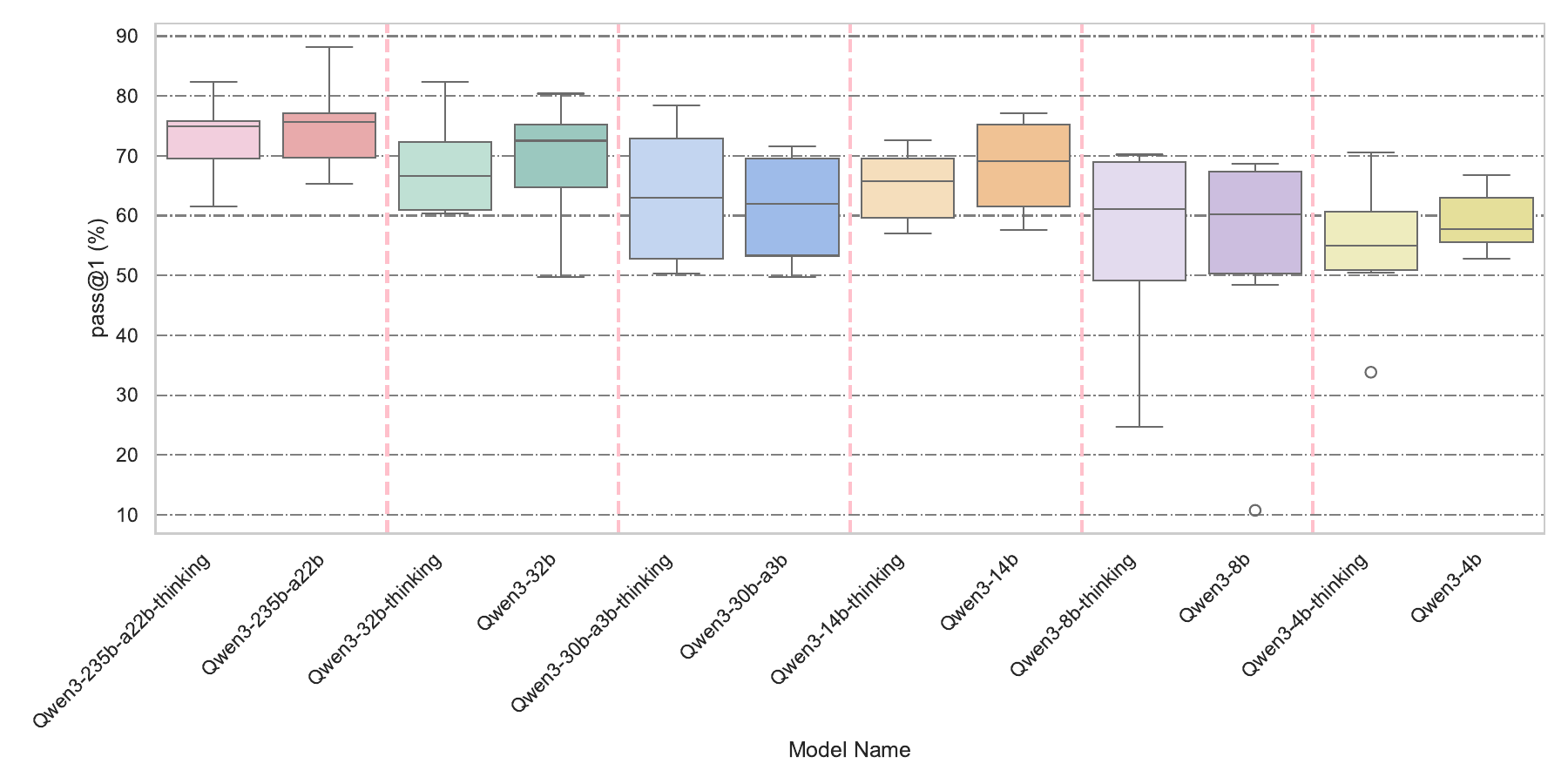}
    \caption{Boxplot of pass@1 performance of Qwen3 series models across programming languages.}
    \label{fig:1a}
  \end{subfigure}
  \hfill
  \begin{subfigure}[b]{0.49\textwidth}
    \includegraphics[width=\linewidth]{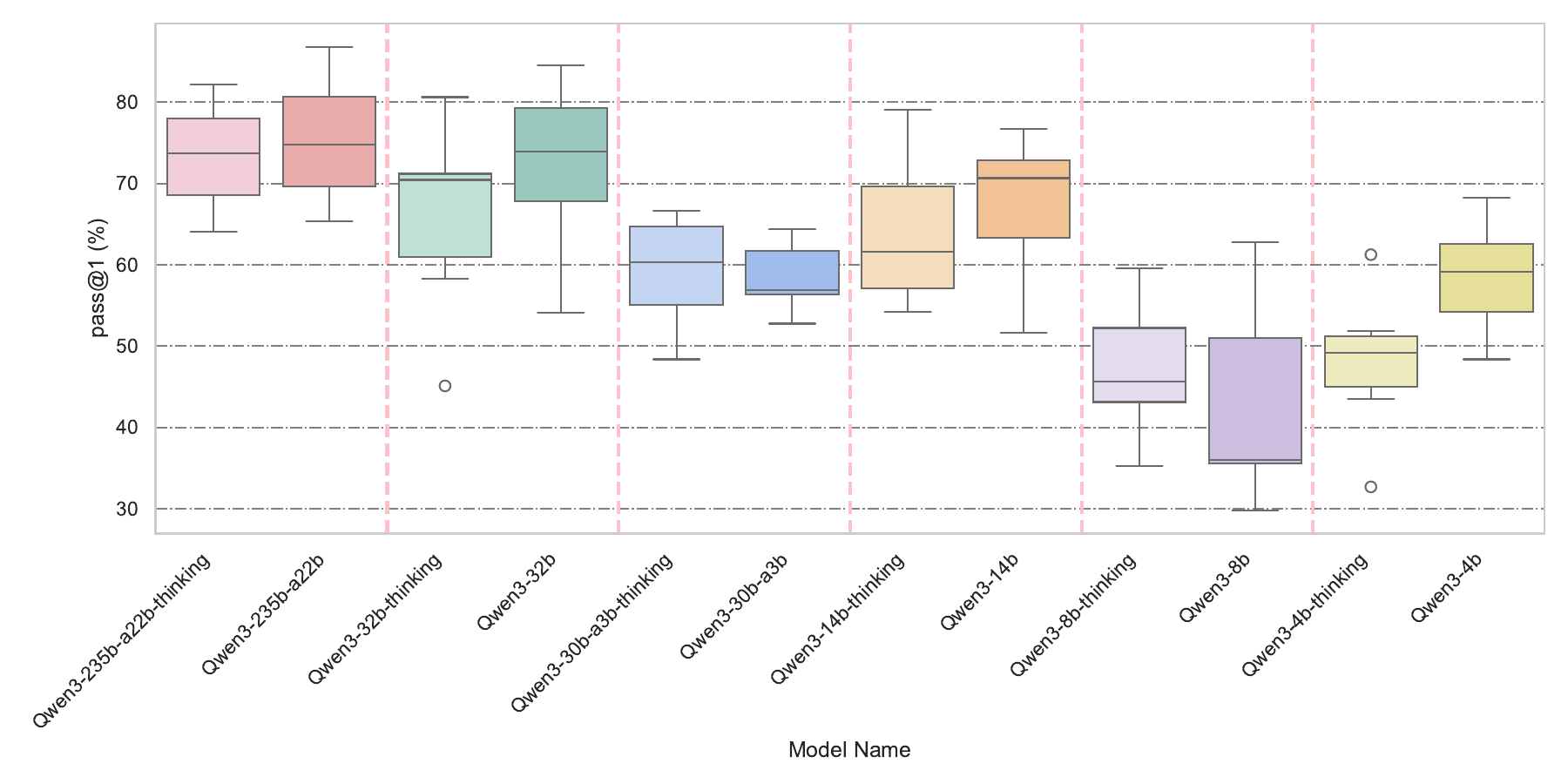}
    \caption{Boxplot of pass@1 performance of Qwen3 series models across IPC techniques.}
    \label{fig:1b}
  \end{subfigure}
  \caption{Boxplot of pass@1 performance of Qwen3 series models.}
  \label{fig:box}
\end{figure*}

We select Qwen3 model family for analysis due to three key reasons: (1) sate-of-the-art performance and latest generation of the family; (2) it offers variants across multiple parameter scales; and (3) it introduces a newly thinking mode.

\textbf{Deviation from scaling law.} Analysis of Qwen3 models (Table~\ref{tab:qwen3_performance} and \ref{tab:qwen3_performance2}) with varying parameter sizes reveals performance on IPC code generation does not consistently improve with larger models. For instance, on Python-based IPC tasks, the smaller Qwen3-4b surpasses larger variants such as Qwen3-8b, 14b, and 30b-A3b in both pass@1 and pass@5. 

\textbf{Impact of thinking mode on IPC code generation.} As shown in Fig.~\ref{fig:box}, incorporating thinking mode in Qwen3 models does not necessarily yield improvements. In fact, performance sometimes declines, suggesting that thinking mode—designed to simulate general-purpose reasoning—may not align well with the structured and protocol-driven characteristics of IPC code. For example, Table~\ref{tab:qwen3_performance} shows that on the C++ subset, the best-performing model without thinking mode achieves a pass@1 of 88.24\%, compared to 82.35\% with thinking mode. A similar drop is observed on the gRPC subset, where pass@1 falls from 86.82\% to 82.17\%. 

\begin{figure*}[t]
  \centering
  \includegraphics[width=0.96\textwidth]{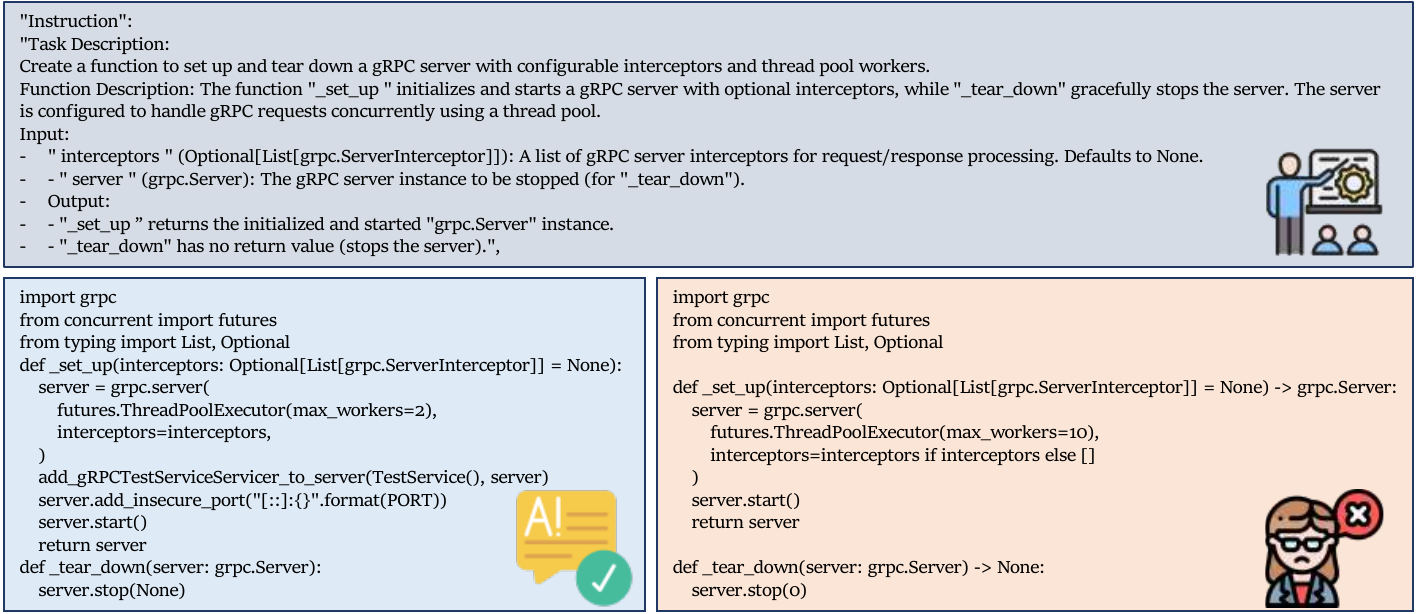}
  \caption{Case Analysis: the 495-th instance in the benchmark.}
  \label{fig:demo}
\end{figure*}
\begin{tcolorbox}[title=Answer to RQ3, boxrule=1pt, boxsep=1pt, left=1pt, right=1pt, top=1pt, bottom=1pt]
Model scale and thinking mode may not necessarily improve CPL code generation. Smaller Qwen3 models sometimes outperform the larger ones. Moreover, thinking mode often reduces the performance for Qwen3, indicating 
that such general-purpose reasoning may not align well with the structured and protocol-driven characteristics of IPC code.
\end{tcolorbox}

\subsection{Case Analysis}

Fig. \ref{fig:demo} illustrates a sample alongside its corresponding ``Instruction'', ``Canonical result'', and an incorrect answer generated by DeepSeek-V3. The task involves implementing a \textit{gRPC} server using the grpcio library. Based on the official documentation, we have summarized the general procedure for such problems, represented as the semantic descriptions of the various states in the FSM: (1) Import the necessary libraries required for server creation; (2) Create a \textit{gRPC} server with a thread pool executor to support concurrency; (3) Bind the server to a port using the \textit{add\_insecure\_port} method; (4) Start the server to begin listening for incoming \textit{gRPC} requests.

From the ``Canonical result'' shown on the left side of Fig. \ref{fig:demo}, the implementation fully complies with this standard and passes validation by the predefined FSM. In contrast, the incorrect answer on the right side lacks the critical step of binding the server to a port using \textit{add\_insecure\_port}. Consequently, the server does not listen on any port, rendering it incapable of receiving requests and thus functionally ineffective.

Notably, although the instruction does not explicitly mention the binding operation, it states: ``Initialize and start a \textit{gRPC} server capable of handling concurrent \textit{gRPC} requests.'' This implicitly requires binding the \textit{gRPC} server to a port. In practice, software engineers seldom specify every detail, as enumerating all functional steps often takes longer than implementing the functionality itself.

\section{Limitations and future work}
\textbf{Limitations.} This paper presents a benchmark for evaluating LLM performance in CPL scenarios. We conducted a comprehensive analysis of CPL interactions in open-source MPL  projects on GitHub, and designed 156 FSMs to locate interaction entry points, validate extracted interaction segments, and assess the correctness of code generated by LLMs. In addition, we developed a workflow powered by LLMs to automate the benchmark construction process. Finally, we evaluated 20 LLMs on this benchmark and derived several insightful findings.

While this is the first benchmark tailored for assessing LLMs in CPL code generation, it has limitations: (1) The distribution of programming languages and IPC techniques in open-source projects is uneven, leading to limited samples in some sub-tasks; (2) Due to the complexity of FFI—such as type mapping, platform-specific setup, and runtime linking—it is not yet included in the benchmark; (3)In the RQ3 experiment, we focused exclusively on the Qwen3 model series. Due to space constraints, we did not extend similar analyses to other model families. Additionally, to align with experimental settings in related work, we did not explore the impact of hyperparameter configurations on performance.

\textbf{Future Work.} (1) Improve LLMs’ ability to generate CPL interaction code through targeted adaptation; (2) Expand benchmark coverage to include more CPL techniques and tasks for broader evaluation.
(3) Conduct more extensive experimental analyses, incorporating a broader range of parameter variations to assess their impact on LLM performance.

\section{Conclusion}
CPL interaction code is prevalent in MPL projects and is essential for their construction. However, current LLMs often fail to generate such code accurately, with performance being highly sensitive to both programming languages and interaction techniques. In addition, the model's thinking mode provides limited benefit and may negatively impact CPL code generation. Enhancing the ability of LLMs to generate CPL interaction code will be critical for scaling up their application to the generation of large-scale MPL software systems in future work.

\bibliographystyle{IEEEtran}
\bibliography{ref}
\end{document}